\documentstyle[aps,epsfig,epsf,multicol]{revtex}
\def\be{\begin{equation}}
\def\ee{\end{equation}}
\def\bearr{\begin{eqnarray}}
\def\eearr{\end{eqnarray}}
\def\zbf#1{{\bf {#1}}}
\def\bfm#1{\mbox{\boldmath $#1$}}
\def\hf{\frac{1}{2}}

\begin{document}
\draft
\preprint{}

\title{Chiral symmetry breaking, color superconductivity
and quark matter phase diagram: a variational approach}
\author{
Hiranmaya Mishra \footnote {email address hm@prl.ernet.in}
and Jitendra C. Parikh \footnote {email address parikh@prl.ernet.in}} 
\address{Theory Division, 
Physical Research Laboratory, Navrangpura, Ahmedabad 380 009, India}
\maketitle
\begin{abstract}
We discuss in this note simultaneous existence of chiral symmetry 
breaking and color superconductivity
at finite temperature and density in a Nambu-Jona-Lasinio type model.
The methodology involves an explicit construction of a variational
ground state and minimisation of the thermodynamic potential.
There exists solutions to the gap equations at finite densities
with both quark antiquark as well as diquark condensates for the ``ground" 
state. However, such a phase is thermodynamically unstable with the pressure
being negative in this region.
We also compute the equation of state, and obtain
the structure of the phase diagram in the model.
\end{abstract}
\vskip 0.5cm
 \section{Introduction}
 The structure of vacuum in
Quantum Chromodynamics (QCD) is one of the most interesting
questions in strong interaction physics \cite{sur}. The evidence for
quark and gluon condensates in vacuum is a reflection of its complex nature
\cite{svz}, whereas chiral symmetry breaking is an essential feature 
 in the description of the low mass hadron properties. Due to the 
nonperturbative nature of QCD in this regime different effective models
have been used to understand the nature of chiral symmetry breaking. 
These have been constructed, for the most part, in the framework of 
a Nambu-Jona- Lasinio (NJL) model with a four fermion interaction.
There also have been attempts to generalise this in the case of
Coulomb gauge QCD with an effective propagator simulating the effects of
confining potentials \cite{chrl,finger,davis,alkofer,klev,bhalerao,amcrl}.
Studies at finite temperatures using imaginary time formulation 
of finite temperature field theory\cite{alkz} have also been carried out. 

Recently there has been a lot of interest in strongly interacting 
matter at high densities.
In particular, a color superconducting phase for it involving
diquark condensates has been considered with a
gap of about 100 MeV . The studies have been done with an effective 
four fermion interaction between quarks \cite{wil}, direct instanton approach 
\cite{sursc} or a perturbative QCD calculation at finite density 
\cite{pisarski}. There has also been
a study of this phase in NJL model \cite{sarah}. In the NJL model
however,the aspect of chiral symmetry breaking in presence of 
diquark condensates has not been considered. Such a question
has been considered in Ref.\cite{berges} in an instanton 
induced four fermion interaction model within a mean field approximation.
In this note, we use a different method to study the problem.
We consider a variational approach with an explicit assumption for
the ground state having both quark antiquark and diquark condensates. The
actual calculations are carried out for the NJL model such that the
minimisation of the free energy density determines which condensate will
exist at what density.

 We organise this paper as follows. In section {\bf II} we construct the 
ground state as a vacuum realignment with quark antiquark as well as 
diquark condensates.
The finite temperature and density effects are included within the framework of
thermofield dynamics \cite{tfd} through a thermal Bogoliubov transformation
involving doubling of the Hilbert space. In section {\bf III} we 
shall consider an effective model as is considered for chiral symmetry breaking
in Coulomb gauge QCD \cite{chrl,finger,davis}. 
To solve the gap equation exactly we shall
further take a four fermion point interaction limit of the same similar to
NJL model. In section {\bf IV} we solve the gap equations for the mass
and the superconducting gap,determine the phase diagram and discuss 
the results. Finally, 
we give a summary in section {\bf V}.

\section{ An ansatz for the ground state} As noted earlier we shall 
 include here the effects of both chiral symmetry breaking as well as
diquark pairing. 
For the consideration of chiral symmetry breaking, we shall take the
perturbative vacuum state with chiral symmetry as $|0\rangle$. In this basis
we shall take the quarks as massless. We shall then
assume a specific vacuum realignment which breaks chiral symmetry
because of interaction.

We have seen earlier that chiral symmetry breaking takes place with the
formation of quark antiquark condensates in perturbative vacuum
\cite{amcrl,hmnj,spmtlk,amspm,spm78,lopr,sinp}. We consider the quark 
field operator
expansion in momentum space given as
 \cite{amcrl,hmnj,spmtlk,amspm}
\begin{eqnarray}
\psi (\zbf x )\equiv &&\frac{1}{(2\pi)^{3/2}}\int \tilde\psi(\zbf k)
e^{i\zbf k\cdot\zbf x}d\zbf k \nonumber\\ 
=&&\frac{1}{(2\pi)^{3/2}}\int \left[U_0(\zbf k)q^0_I(\zbf k )
+V_0(-\zbf k)\tilde q^0_I(-\zbf k )\right]e^{i\zbf k\cdot \zbf x}d \zbf k,
\label{psiexp}
\end{eqnarray}
where \cite{spm78,lopr}
\begin{eqnarray}
U_0(\zbf k )=&&\frac{1}{\sqrt{2}}\left(\begin{array}{c}1\\
\zbf \sigma \cdot \hat k \end{array}\right),
\nonumber\\
V_0(-\zbf k )=&&\frac{1}{\sqrt{2}}
\left( \begin{array}{c} -\zbf \sigma \cdot \hat k 
\\ 1\end{array}
\right).
\label{uv0}
\end{eqnarray}
The subscript $I$ and the superscript $0$ indicate that the operators 
$q_I^0$ and $\tilde q_I^0$ are
two component ones which annihilate or create quanta acting upon
 the perturbative or the chiral vacuum.

We now consider vacuum destabilisation leading to chiral symmetry breaking
\cite{amcrl,hmnj,spmtlk,amspm} described by,
\begin{equation} 
|vac\rangle={\cal U}_Q|0\rangle,
\label{u0}
\end{equation} 
where
\begin{equation}
{\cal U}_Q=\exp\left (
\int q_I^0(\zbf k)^\dagger(\bfm {\sigma }\cdot\zbf k) h(\zbf k)
 \tilde q_I^{0} (\zbf k)d\zbf k-h.c.\right ).
\label{u1}
\end{equation}
 
In the above,  we have suppresed the flavor and color indices on the 
two component quark and antiquark operators. Further $h(\zbf k)$  is a
real function of $|\zbf k|$ which describes vacuum realignment for
quarks of a given flavor. We 
consider here two flavors and take
the condensate function $h(\zbf k)$ to be the same for u and d quarks.
Clearly, a nontrivial $h(\zbf k)$ shall break chiral
$SU(2)_L\times SU(2)_R$ symmetry to the custodial symmetry $SU(2)_V$
for the light quark doublet \cite{amspm,sinp}. 
In what follows we shall exploit this result.

Having defined the state as in Eq.(\ref{u0}) for chiral symmetry breaking,
we shall next define the state involving diquarks. We note that
as per BCS result such a state will be dynamically favored
if there is an  attractive interaction between the quarks \cite{bcs}.
Such an interaction exists in QCD in the qq color antitriplet, Lorentz 
scalar and isospin singlet channel. In the flavor antisymmetric channel
 the interaction can be scalar, pseudoscalar or vector whereas in 
flavor symmetric channel only the axial vector channel is attractive.
In the present work, we shall consider the ansatz
 state involving diquarks as 
\begin{equation}
|\Omega\rangle=U_d|vac\rangle=\exp(B_d^\dagger-B_d)|vac\rangle,
\label{omg}
\end{equation}
where
\begin{equation}
{\cal B}_d ^\dagger=\frac{1}{2}\int \left[q_r^{ia}(\zbf k)^\dagger
f(\zbf k) q_{-r}^{jb}(-\zbf k)^\dagger
\epsilon_{ij}\epsilon_{3ab}
+\tilde q_r^{ia}(\zbf k)
f_1(\zbf k) \tilde q_{-r}^{jb}(-\zbf k)
\epsilon_{ij}\epsilon_{3ab}\right]
d\zbf k.
\label{bd}
\end{equation}
\noindent 
In the above, $i,j$ are flavor indices, $a,b$ are the
color indices and $r(=\pm 1/2) $ is the spin index. As noted earlier we 
shall consider systems with two flavors and three colors. 
 We have also introduced  here two trial functions $f(\zbf{k})$ and
$f_1(\zbf k)$ respectively for the diquark and diantiquark
channel. As may be noted the state constructed in eq.(\ref{omg}) is spin
singlet and is antisymmetric in color and flavor. The corresponding Bogoliubov
transformation for the operators is given by

\begin{eqnarray}
\left(
\begin{array}{c}q_{Ir}^{ia\prime}(\zbf k)\\ 
 q_{I-r}^{kc\prime}(-\zbf k)^\dagger\end{array} 
\right)& =&
\left(
\begin{array}{cc} \cos {f(\zbf k)}  &
 -r\frac{f(\zbf k)}{|f(zbf k)|}\epsilon_{ik}\epsilon_{3ac}\sin f(\zbf{k})\\
 r\frac{f^\star(\zbf k)}{|f(\zbf k)|}\epsilon_{ki}\epsilon_{3ca}\sin f(\zbf{k})
 & \cos{f(\zbf k)}
  \end{array}
 \right)
\left(
\begin{array}{c}q_{Ir}^{ia}(\zbf k)\\ 
\tilde q_{I-r}^{kc}(-\zbf k)\end{array} 
\right).
\label{uquqpd}
\end{eqnarray}
In a similar manner one can write down the Bogoliubov transformation for 
the antiquark operators corresponding to $|\Omega\rangle$ basis.

Finally, to include the effect of temperature and density we obtain the 
state at finite temperature and density $|\Omega(\beta,\mu)\rangle$ by 
a thermal Bogoliubov transformation over the state $|\Omega\rangle$ 
using thermofield dynamics (TFD) as described in ref.s \cite{tfd,amph4,cort} .
We have,
\begin{equation} 
|\Omega(\beta,\mu)\rangle={\cal U}_{\beta,\mu}|\Omega\rangle
\label{ubt}
\end{equation} 
where ${\cal U}_{\beta,\mu}$ is
\begin{equation}
{\cal U}_{\beta,\mu}=e^{{\cal B}^{\dagger}(\beta,\mu)-{\cal B}(\beta,\mu)},
\label{ubm}
\end{equation}
with, 
\begin{equation}
{\cal B}^\dagger(\beta,\mu)=\int 
q_I^\prime (\zbf k)^\dagger \theta_-(\zbf k, \beta,\mu)
\underline q_I^{\prime} (\zbf k)^\dagger +
\tilde q_I^\prime (\zbf k) \theta_+(\zbf k, \beta,\mu)
\underline { \tilde q}_I^{\prime} (\zbf k) d\zbf k.
\label{bth}
\end{equation}
In Eq.(\ref{bth}) the ansatz functions $\theta_{\pm}(\zbf k,\beta,\mu)$
will be related to quark and antiquark distributions and the underlined
operators are the operators in the extended Hilbert space associated with
thermal doubling in TFD method.
 Thus the ansatz functions at finite temperature and density given 
in Eq.(\ref{ubt}) invloves five functions - $h(\zbf k)$,
for the quark anti quark condensates, $f(\zbf k)$ and $f_1(\zbf k)$ describing
respectively the diquark and diantiquark condensates and 
$\theta_{\pm}(\zbf k,\beta,\mu)$ to include the temperature and 
density effects. All these functions are to be obtained by minimising the
thermodynamic potential. This will involve an 
assumption about the effective 
hamiltonian . We shall carry out this minimisation
 in the next section.

\section{estimation of the pressure and the model}
Since the low and medium energy behaviour of QCD is not well understood one has to make assumptions about the effective interaction between the quarks. To
consider chiral symmetry breaking in the Coulomb gauge pairing model
\cite{finger,davis,alkofer} one had the effective action
\be
{\cal {S}}_{eff}=\int d^4x \bar \psi (x) (i{\gamma^\mu{\partial}_\mu}) \psi (x)
+\frac{1}{2}
\int d^4x\int d^4y g^2(x-y)J_{\mu}^a D_{ab}^{\mu\nu}(x-y)J_\nu^b(y),
\label{seff}
\end{equation}
where,
$$J_\mu^a(x)=\bar \psi^k(x)\gamma_\mu (\frac{\lambda^a}{2}) \psi^k(x).$$ 
A sum over color indices $a$ and $b$ is implied, $g$ is the strong coupling
 constant and,
$D^{\mu\nu}_{ab}(x-y)$ is the full gluon propagator. 
This effective action may be thought of as resulting from integrating out the 
gluonic degrees of freedom from the full QCD Lagrangian keeping only 
the bilinear terms of the quark currents. While dealing with the gluon sector, 
to make things computable, one uses different approximations for the full
propagator which is written down in Coulomb gauge
\cite{finger,davis,alkofer,amcrl,hmnj}. The most obvious one is to
 neglect transeverse gluons
completely maintaining only the instantaneous Coulomb interactions.
This was done in Ref.\cite{finger,davis} for chiral symmetry breaking.
Another possibilty is to neglect retardation effects for the transverse
gluons and have e.g. $D^{ij}(p)=D^{ij}(\zbf p,p_0)$ - the so called Breit 
interaction. In such instantaneous approximations one can interpret the
product of the coupling constant and the gluon propagator as the Fourier
tranform of the effective quark antiquark potential.
Different effective potentials such as confining, Richardson, and
a screened potential for the transverse part have been considered for
chiral symmetry breaking in Coulomb gauge pairing model\cite
{finger,davis,alkofer}.

To deal with the case of finite temperature and density one might take
one loop resummed perturbative QCD results at finite temperature and densities
\cite{pisarski,leblac} for the propagator. However, we shall here assume
a point like interaction approximation. The reason we shall choose this 
is its simplicity regarding its solvability.
Further we can compare the results with those of NJL 
model calculations \cite{sarah}.
The delta function interaction produces short distance singularities and so to regulate the integrals we shall restrict the phase space inside the sphere
$|\zbf p|< \Lambda$. Thus the effective Hamiltonian we shall be considering
is given by
\be
{\cal H}=\psi^\dagger(-i\bfm \alpha \cdot \bfm \nabla )\psi
+\frac{g^2}{2}J_\mu^aJ^{\mu a}.
\label{ham}
\ee

We next write down the expectation values of various 
operators in the thermal vacuum given in Eq.(\ref{ubt}).
These expressions would be used to calculate thermal expectation value of the
Hamiltonian to compute the thermodynamic potential.
We define
\begin{equation}
\langle \Omega(\beta,\mu)
 |\tilde\psi_\alpha(\zbf k)\tilde\psi_\beta(\zbf k')^\dagger
|\Omega(\beta,\mu)\rangle
=\Lambda_{+\alpha\beta}(\zbf k,\beta,\mu)\delta(\zbf k-\zbf k'),
\label{psipsidb}
\end{equation}
and,
\begin{equation}
\langle \Omega(\beta,\mu)
|\tilde\psi_\beta(\zbf k)^\dagger\tilde\psi_\alpha(\zbf k')
|\Omega(\beta,\mu)\rangle
=\Lambda_{-\alpha\beta}(\zbf k,\beta,\mu)\delta(\zbf k-\zbf k'),
\label{psidpsib}
\end{equation}
where,
\begin{equation}
\Lambda_+(\zbf k,\beta,\mu)
=\hf\left(1+F_1(\zbf k)
-F(\zbf k)+\big(\gamma^0\sin 2h(\zbf k)+\bfm\alpha\cdot\hat\zbf k\cos 2h
(\zbf k)\big)\big(1-F(\zbf k)-F_1(\zbf k)\big)
\right),
\label{prpb}
\end{equation}
and,
\begin{equation}
\Lambda_-(\zbf k,\beta,\mu)
=\hf\left(1+F(\zbf k)-F_1(\zbf k)-\big(\gamma^0\sin 2h(\zbf k)
+\bfm\alpha\cdot\hat\zbf k\cos2 h (\zbf k)\big)\big(1-F(\zbf k)-
F_1(\zbf k)\big)
\right).
\label{prmb}
\end{equation}
Here, the effect of diquark condensate and the temperature and/or density
dependence is encoded in the function $F(\zbf k)$ and $F_1(\zbf k)$ given as
\begin{equation}
F(\zbf k)=\sin^2 f(\zbf k)+\cos 2f(\zbf k) \sin^2\theta_-(\zbf k,\beta,\mu),
\label{fkb}
\end{equation}
and,
\begin{equation}
F_1(\zbf k)=\sin^2 f_1(\zbf k)+\cos 2f_1(\zbf k) \sin^2\theta_+(\zbf k,
\beta,\mu).
\label{f1kb}
\end{equation}
Clearly, at zero temperature and zero density the functions $F$ and $F_1$
vanish and the projection operators reduce to the forms
considered earlier\cite{hmnj},
\be
\Lambda_{\pm}(\vec k, \beta)=\frac{1}{2}\bigg[ 
1\pm (\gamma^0 \sin\!2h(k) + \vec\alpha\cdot\hat k 
\cos\!2h(k))\bigg].
\label{prpm}
\ee

Further, at zero density but finite temperature $\theta_-=\theta_+$ and
projection operators reduce to \cite{cort},
\be
\Lambda_{\pm}(\vec k, \beta)=\frac{1}{2}\bigg[ 
1\pm \cos 2\theta(\gamma^0 \sin\!2h(k) + \vec\alpha\cdot\hat k 
\cos\!2h(k))\bigg].
\label{lpm}
\ee
\noindent
We also have

\bearr
\langle \Omega(\beta,\mu)| \psi^{ia}_\alpha(\vec x)\psi^{jb}_\gamma
|\Omega(\beta,\mu) \rangle
&=&\frac{\epsilon^{ij}\epsilon^{3ab}}{(2\pi)^3}\int e^{i\zbf k\cdot\zbf x}
{\cal {P}}_{+\gamma\alpha}(\zbf k,\beta,\mu)d\zbf k,\nonumber\\
\langle \Omega(\beta,\mu)| \psi^{ia\dagger}_\alpha(\vec x)
\psi^{jb\dagger}_\gamma
|\Omega(\beta,\mu) \rangle
&=&\frac{\epsilon^{ij}\epsilon^{3ab}}{(2\pi)^3}\int e^{i\zbf k\cdot\zbf x}
{\cal {P}}_{-\alpha\gamma}(\zbf k,\beta,\mu)d\zbf k,
\label{psi}
\eearr

where, 
\begin{equation}
{\cal{P}}_+(\zbf k,\beta,\mu)
=\frac{1}{4}\left[S(\zbf k)+\left(\gamma^0\sin 2h(\zbf k)-\bfm\alpha\cdot
\hat\zbf k\cos 2h (\zbf k)\right)A(\zbf k)
\right]\gamma_5 C,
\label{calpp}
\end{equation}
and,
\begin{equation}
{\cal{P}}_-(\zbf k,\beta,\mu)
=\frac{C\gamma_5}{4}\left[(S(\zbf k)+\left(\gamma^0\sin 2h(\zbf k)
-\bfm\alpha\cdot \hat\zbf k\cos 2h (\zbf k)\right)A(\zbf k)
\right].
\label{calpm}
\end{equation}

\noindent Here, $C=i\gamma^2 \gamma^0$ is the charge conjugation matrix (we
use the notation of Bjorken and Drell) and the functions $S(\zbf k)$ and 
$A(\zbf k)$ are given as,
\begin{mathletters}
\be
S(\zbf k)=\sin\!2f(\zbf k)\cos 2\theta_-(\zbf k,\beta,\mu)
+\sin\!2f_1(\zbf k)\cos 2\theta_+(\zbf k,\beta,\mu),
\label{sk}
\ee
and,
\begin{equation}
A(\zbf k)=\sin\!2f(\zbf k)\cos 2\theta_-(\zbf k,\beta,\mu)
-\sin\!2f_1(\zbf k)\cos 2\theta_+(\zbf k,\beta,\mu),
\label{ak}
\end{equation}
\end{mathletters}

Using Eq. (\ref{psidpsib}) we have for the kinetic energy part
\be
T\equiv
 \langle \Omega,\beta,\mu|
\psi^\dagger
(-i\bfm \alpha \cdot \bfm \nabla )\psi
| \Omega\beta\mu\rangle
=\frac{\gamma}{(2\pi)^3}\int d \zbf k \cos 2h(\zbf k)(1-F-F_1),
\label{t}
\ee
where, the degenracy factor $\gamma=12$ corresponding to spin(2), color(3) and
flavor(2) degrees of freedom.
$T$ in Eq.(\ref{t}) above, however, includes the contribution of the 
perturbative zero point energy corresponding to $h(\zbf k)=0=f(\zbf k)=f_1(\zbf k)$.
Subtracting this out we have the contribution from the kinetic energy
\be
{\cal {T}}=T-\langle \hat T \rangle|_{h=f=f_1=\theta_{\pm}=0}
=\frac{\gamma}{(2\pi)^3} \int |\zbf k|
\left [2 \sin^2 h(\zbf k)+\cos 2 h(\zbf k) 
(F+F_1)\right ].
\label {tren}
\ee

Similarly the contribution from the interaction term in Eq.(\ref{ham})
after subtracting out the zero point pertuturbative energy turns out to be
\begin{equation}
{\cal {V}}\equiv
 \langle \Omega,\beta,\mu|\frac{g^2}{2}J_\mu^aJ^{\mu a}
| \Omega\beta\mu\rangle
=V_1+ V_2
\label{v}
\ee
where, the contribution $V_1$ arises from using Eq.s (\ref{psipsidb}),
(\ref{psidpsib})
and is given as
\be
V_1=\frac{g^2 }{2}N_f \frac{(N_c^2-1)}{2} 2(I_1^2-2 I_2^2)=8g^2(I_1^2-2 I_2^2)
\label{v1}
\ee
with,
\be
I_1=\frac{1}{(2\pi)^3}\int d \zbf k(F-F_1)
\label{i1}
\ee
and
\be
I_2=\frac{1}{(2\pi)^3}\int d \zbf k(1-F-F_1)\sin 2h(\zbf k).
\label{i2}
\ee

The term $V_2$ arises from using Eq.(\ref{calpp})and (\ref{calpm})
and we have
\be
V_2=\frac{2g^2}{3}(-2I_3^2+I_4^2),
\label{v2}
\ee
where
\be
I_3=\frac{1}{(2\pi)^3}\int d \zbf k S(\zbf k)
\label{i3}
\ee
and

\be
I_4=\frac{1}{(2\pi)^3}\int d \zbf k A(\zbf k)\sin 2 h(\zbf k).
\label{i4}
\ee
Then the free energy density is
\be
{\cal F}=\epsilon -\mu N={\cal T}+{\cal V} -\mu N
\label{calf}
\ee
where,  $\mu$ is the chemical potential and $N$ is  quark number density.
 Further,
\be
N=\langle \psi^{\dagger} \psi \rangle=\frac{\gamma}{(2\pi)^3}
\int d \zbf k (F-F_1)=\gamma I_1.
\label{nden}
\ee
Finally, for the entropy density we have\cite{tfd}
\be
s=-\frac{\gamma}{(2\pi)^3}\int d \zbf k
\left ( \sin^2\theta_-ln \sin^2\theta_-
+\cos^2\theta_-ln \cos^2\theta_-
+sin^2\theta_+ln sin^2\theta_+
+\cos^2\theta_+ln \cos^2\theta_+\right )
\label{ent}
\ee
and the thermodynamic potential which is negative of pressure is given by
\cite{qcdtb}
\be
{\Omega}=-{\cal P}=\epsilon-\mu N-\frac{1}{\beta} s
\label{pres}
\ee
Now if we minimise the thermodynamic potential $\Omega$ with respect to
$h(\zbf k)$,we get
\be
tan 2 h(\zbf k)=\frac{8g^2I_2}{3|\zbf k|}\equiv \frac{M}{|\zbf k|}.
\label{tan2h}
\ee
where, $M=8 {g^2 I_2}/{3}$.
Substituting this back in Eq.(\ref{i2}) we have the mass gap equation 

\be
M=\frac{8 g^2I_2}{3}=\frac{8 g^2}{3}
\frac{1}{(2\pi)^3}\int \frac{M}{\sqrt{\zbf k^2 +M^2}} (1-F-F_1) d
\zbf k.
\label{mgap}
\ee
Clearly, the above includes the effect of diquark condensates 
as well as temperature and density through the functions $F$ and $F_1$ 
given in Eq.s (\ref{fkb})
and (\ref{f1kb}) respectively.
 The zero temperature
and density limit is given by setting $F=0=F_1$ which has the same structure
as in NJL model \cite{klev,sarah,hmnj}.

Next, minimising the thermodynamic potential $\Omega$ with respect to
$f$ and $f_1$ yields
\be
tan 2f(\zbf k)=\frac{2g^2}{9}
\frac{\left(2I_3-I_4 \sin 2h\right)}
{\sqrt{\zbf k^2+M^2} -\left(\mu-\frac{4g^2}{3}I_1\right )}
\label{tan2f}
\ee
and,
\be
tan 2f_1(\zbf k)=\frac{2g^2}{9}\frac{\left(2I_3+ I_4\sin 2h\right)}
{\sqrt{\zbf k^2+M^2} +\left(\mu-\frac{4g^2}{3}I_1\right )}.
\label{tan2f1}
\ee
We might further simplify equations (\ref{tan2f}) and (\ref{tan2f1}) by
noting from Eq.s (\ref{i3}), (\ref{i4}), (\ref{ak}) and (\ref{sk})
that the integral $I_4$ is small compared to $I_3$, as the integrand 
in the former is a difference of two postive quantities whereas the latter is a 
sum of these two quantities. In addition, the integrand in  $I_4$ is
suppresed by the quark antiquark condensate function $\sin 2h(\zbf k)$ 
which is not the case in $I_3$. Finally, this apart, in Eq.s (\ref{tan2f})
and (\ref{tan2f1}) the numerical coefficient in the second term of the
numerator is small compared to that in the first term. Thus, we can approximate
\be
tan 2f(\zbf k)=\frac{\Delta}{E-\nu}
\label{df}
\ee
\be
tan 2f_1(\zbf k)=\frac{\Delta}{E+\nu},
\label{df1}
\ee
where, the superconducting gap $\Delta=(4
g^2/9) I_3$, $E=\sqrt{\zbf k^2 +M^2}$ and  $\nu=\mu-4g^2/3 I_1$ is the chemical
potential in presence of interaction \cite{wil}. From the definition of 
superconducting gap and Eq.(\ref{i3}) we have the superconducting gap equation 
given by
\be
\Delta=\frac{4g^2}{9}I_3=\frac{4g^2}{9(2\pi)^3}\int d \zbf k
\left(\frac{\Delta}{\sqrt{\Delta^2+(E-\nu)^2}}\cos 2\theta_-(\zbf k,\beta,\mu)
+\frac{\Delta}{\sqrt{\Delta^2+(E+\nu)^2}}\cos 2\theta_+(\zbf k,\beta,\mu)
\right )
\label{scgap}
\ee
Finally, minimisation of the thermodynamic potential with respect to the
thermal functions $\theta_{\pm}(\zbf k,\beta,\mu)$ gives
\be
\sin^2\theta_-=\frac{1}{\exp(\beta\omega_-)+1}
\label{them}
\ee
and
\be
\sin^2\theta_+=\frac{1}{\exp(\beta\omega_+)+1}
\label{thep}
\ee
where, $\omega_{\pm}=\sqrt{\Delta^2+\xi_{\pm}^2}$ and $\xi_{\pm}=(E\pm\nu)$.
Therefore, with all the ansatz functions {\it determined}  from the 
minimisation of the thermodynamic potential ${\Omega(\beta,\mu)}$,
the mass gap equation (\ref{mgap})
and superconducting gap equations are given respectively as
\be
\frac{4g^2}{3}\frac{1} {(2\pi)^3}
\int d\zbf k \frac{1}{\sqrt{\zbf k^2+M^2}}
\left(\frac{\xi_-}{\omega_-}tan h(\frac{\beta\omega_-}{2}) 
+\frac{\xi_+}{\omega_+}tan h(\frac{\beta\omega_+}{2})
\right)=1
\label{mgap1}
\ee 
\be
\frac{4g^2}{9}\frac{1} {(2\pi)^3}
\int d\zbf k
\left(\frac{tan h(\frac{\beta\omega_-}{2})}{\omega_-}
+ \frac{tan h(\frac{\beta\omega_+}{2})}{\omega_+}
\right)=1
\label{scgap1}
\ee 

Equation (\ref{mgap1}) is the generalisation of the mass gap equation
considered at zero temperature and density of Ref.\cite{klev,hmnj} to
include the effect of finite temperature and density \cite{sarah}
{\it alongwith} the effect of diquark condensates.
Similarly, Eq.(\ref{scgap1}) is the relativistic generalisation
of the BCS gap equation \cite{sarah} in the presence of a dynamically generated
mass gap through a quark antiquark condensate structure for the vacuum.
These are the main new features of the present work.

\section{Solution of the gap equations and results}

Before trying to solve the coupled gap equations (\ref{mgap1}) and
(\ref{scgap1}) we discuss the solutions in the following two 
different limiting cases, so that we make connection with
earlier studies \cite{wil,sarah}. \\
\begin{itemize}
\item  Case--I :Only quark antiquark condensate i.e. $h(\zbf k)\neq 0;
f(\zbf k)=0=f_1(\zbf k)$\\
In this limit $\Delta=0$ and the gap equation (\ref{mgap1}) reduces to
\be
M=\frac{8 g^2}{3}\frac{1}{(2\pi)^3}\int\frac{M}{\sqrt{\zbf k^2+M^2}}(1-n_+
-n_-) d\zbf k
\label{mgap0}
\ee
where, $$ n_{\pm}=\frac{1}{\exp(\beta(E(\zbf k)\pm\nu))+1}$$
This is the same equation as obtained in Ref.\cite{klev,asakawa,alexander}
except for numerical factors before the integrand. This is due to the
difference
in the Lorentz structure of the four fermion interaction term. In 
Ref. \cite{klev} the interaction term is $(\bar\psi\psi)^2
+((\bar\psi\gamma_5\tau
\psi)^2$ that can originate from a Fierz transformation of the
$J_\mu^a J^{\mu a}$ interaction as considered here. In Ref.\cite{alexander}
only $ J_0^a J^{0 a}$ term was included. We might however mention here that
keeping $J_0J_0$ term only will make the pressure negative at high densities
even when the condensate vanishes. At zero temperature the mass gap is 
given by the solution of the equation
\be
\frac{4 g^2}{3\pi^2}\int_{k_f}^\Lambda\frac{k^2 dk}{\sqrt{k^2+M^2}}=1
\label{mgap00}
\ee
where the fermi mometum $k_f$is defined through the (interacting) chemical
potential $\nu$ as $k_f^2+M^2=\nu^2$. We also note that this gap equation is derived here by minimising the thermodynamic potential over a variational
ansatz and not through a mean field approximation \cite{asakawa} or a Hartree
Fock calculation\cite{klev}. The critical fermi momentum is given 
by the density where $M=0$ and is given at zero temperature by,

\be
k_f^c=\Lambda(1-\frac{3\pi^2}{2 g^2\Lambda^2})^{1/2}
\label{kfc}
\ee
 and
the corresponding quark number density is given by
$n_0={\gamma k_f^{c3}}/{6\pi^2}$. The variation of 
mass gap as a function of fermi momentum
is shown in Fig. (\ref{mgapfig}). We have chosen the value of the coupling
$g^2=55.98 GeV^{-2}$ and $\Lambda=0.67 GeV$ as  typical values.
In fact, they have been chosen so that the diquark condensate function vanishes
at the same critical temperature as in Ref.\cite{wil,sarah}.

\begin{figure}[htbp]
\begin{center}
\epsfig{file=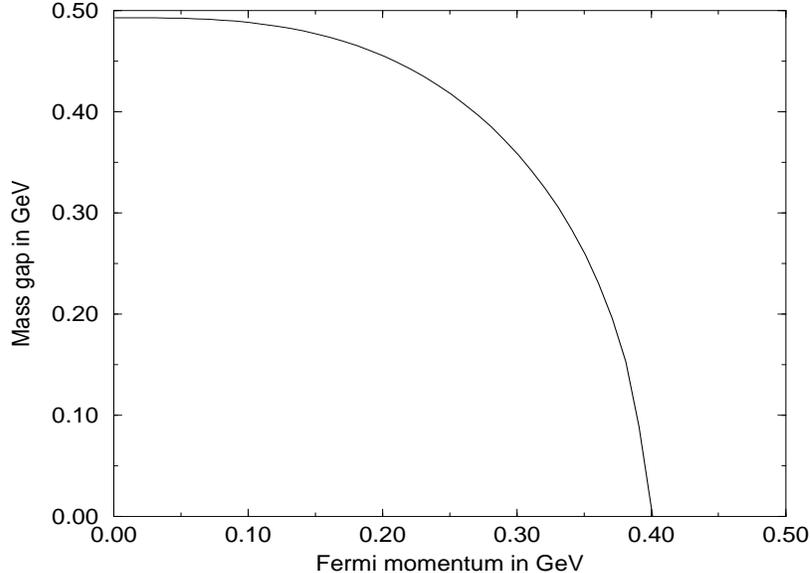,width=12 cm,height=9cm}
\end{center}
\caption{\em Mass gap vs fermi mometum. Critical fermi mometum turns out to be
0.39 GeV corresponding to a quark number density of 1.7 per $fm^3$.}
\label{mgapfig}
\end{figure}

With this choice of parameters, the dynamically generated quark mass
 at zero temperature and density becomes about 490 GeV which is 
rather high compared to the standard value of around 300 MeV.
With a lower value of $g^2\Lambda^2$ the same could be made smaller. 
The critical density in this model turns out to be about $1.7/fm^3$
 corresponding to a value of the fermi momentum $k_f^c$
 of $0.4 GeV$. In Fig.2 we have plotted the
pressure as a function of fermi momentum. Here we have added the bag constant 
$\epsilon_0$ which is the energy density at zero quark number density at zero
temperature \cite{bhalerao}. This turns out to be about $-(0.2 GeV)^4$. As
observed earlier \cite{wil} the pressure has a cusp like structure and becomes
negative at finite density. The portion of the curve that goes down
with fermi momentum corresponds to a nontrivial mass gap solution and the portion that increases with density correponds to zero mass solution of the mass 
gap equation (\ref{mgap1}). As in Ref\cite{sarah}, and unlike Ref.\cite{wil}
the effects of interactions do not disappear beyond the chiral symmetry 
restoration phase. This is due to the fact that there is no running coupling
 involved here and in fact, the pressure grows as $n^2$ or $k_f^6$ at 
high densities as may be expected from the interaction term $V_1$ involving
$I_1^2=N^2/\gamma^2$ given in Eq.(\ref{v1}).
However, one has to keep in mind that to be consistent with the
philosophy of high momentum cutoff $\Lambda$, the fermi momentum 
should be less than the cut off momentum.
The negative pressure at intermediate densities can be understood in terms of
mechanical instability and  can have the interpretation that
uniform nonzero density quark matter will break up into
droplets of finite density in which chiral symmetry is
restored surrounded by empty space with zero pressure and density.
It is tempting to identify these droplets with nucleons
within which the density is nonzero and $\langle \bar q q\rangle$
=0--- a fact reminiscent of bag models. Nothing within the
model however implies that the droplets have quark number three \cite{wil}.
\begin{figure}
\begin{center}
\epsfig{file=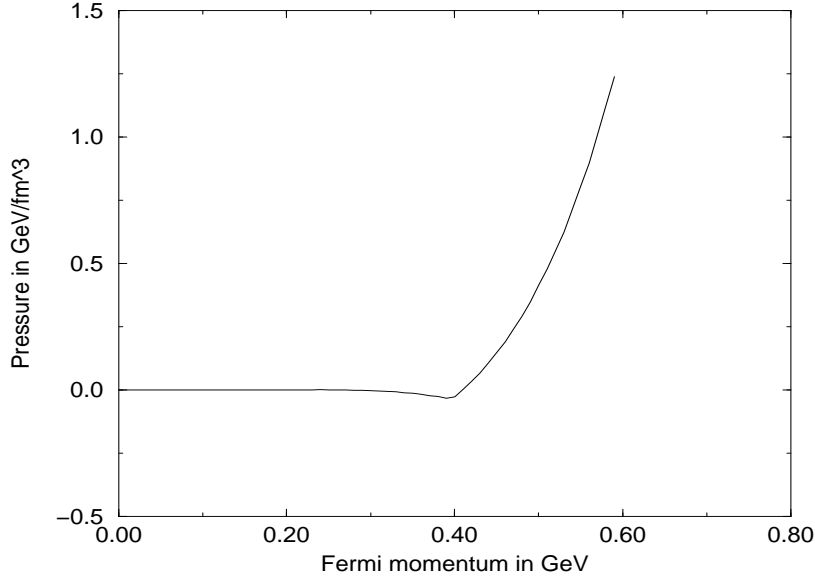,width=12 cm,height=9cm}
\end{center}
\caption{\em Pressure as a function of fermi mometum.
 The decreasing part of the curve
corresponds to $m\neq 0$ solution and the rising part of the curve 
corresponds to $m=0$ solution of the mass gap equation (\ref{mgap1})}
\label{eosfig}
\end{figure}

\item Case-- II:No chiral condensate but only diquark condensate i.e.
$h(\zbf k)=0$ $f(\zbf k)\neq 0$, $f_1(\zbf k)\ne 0$\\ 

In this case, the quark mass arising from chiral condensate is zero and the
superconducting gap equation (\ref{scgap1}) reduces to
\be
1=\frac{2 g^2}{9 \pi^2}\int_0^\Lambda k^2 dk \left[
\frac{1}{\omega_- }tan h(\frac{\beta\omega_-}{2})
+\frac{1}{\omega_+} tan h(\frac{\beta\omega_+}{2})\right]
\label{dscgap}
\ee

where, as before $\omega_{\pm}=\sqrt{\Delta^2+\xi_{\pm}^2}$ but, with
$\xi_{\pm}=|\zbf k|\pm\nu$. This is the relativistic generalisation of
BCS gap equation in superconductivity \cite{bcs}. It has the same
structure as in Ref\cite{wil} in the limit the form factor
is replaced by a sharp cutoff as in Ref.\cite{sarah} and  apart from
the numerical factor preceeding the integrand. Further, this  can be 
rearranged to interpret separately  contributions
arising from particles, antiparticles and holes \cite{wil}. We have chosen the values of the coupling and the cutoff as in Ref.\cite{sarah} so as to get the
same critical temperature as in Ref.\cite{wil}. The gap equation at
zero temperature
is given as
\be
1=\frac{2 g^2}{9 \pi^2}\int_0^\Lambda k^2 dk \left[
\frac{1}{\omega_- }+\frac{1}{\omega_+}\right]
\label{dscgap0}
\ee

The superconducting gap $\Delta$ is plotted in Fig \ref{dscgapfig}.
While obtaining the solution of the gap equation (\ref{dscgap0})
we also insist that a nontrivial solution of the gap equation is acceptable 
only if the corresponding free energy is smaller than that with no gap.         
The gap  increases with fermi momentum,  has a maximum of about 90 MeV
 around fermi momentum 550 MeV, beyond which the effect
of cutoff is felt and it vanishes at around $k_f=600$ MeV.
\begin{figure}
\begin{center}
\epsfig{file=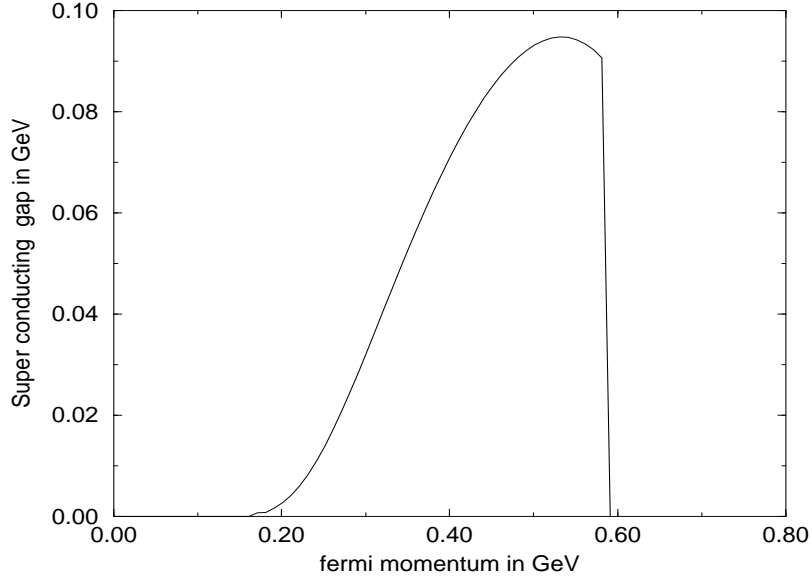,width=12 cm,height=9cm}
\end{center}
\caption{\em Superconducting  gap vs fermi mometum. 
}
\label{dscgapfig}
\end{figure}

The resulting equation of state i.e. pressure as a function of fermi momentum
is plotted in Fig.4. There is no unstable phase as discussed for chiral 
condensate case with negative pressure. Because of the 
interaction term, the equation of state
does not go over to the free massless equation of state
 beyond the critical density. We have
plotted the pressure in Fig.4 for fermi momentum upto 600 MeV noting that the cutoff here is about 650 MeV. 
\begin{figure}
\begin{center}
\epsfig{file=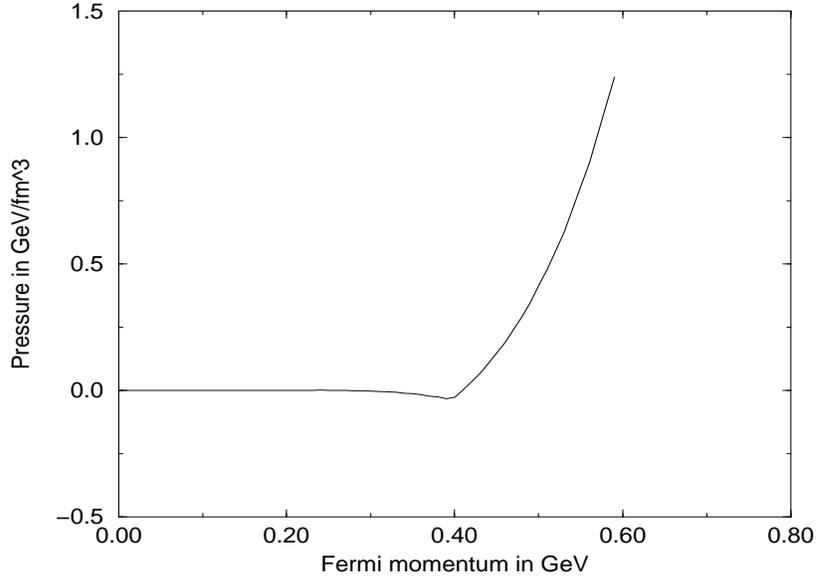,width=12 cm,height=9cm}
\end{center}
\caption{\em Pressure vs fermi mometum. 
}
\label{dprefig}
\end{figure}

\item Case-- III: $h(\zbf k)\neq 0;$$f(\zbf k)\ne 0$and $f_1(\zbf k)\ne 0$.\\
Here we solve the coupled gap equations (\ref{mgap1}) and (\ref{scgap1})
which at zero temperature but finite density reduce to
\begin{equation}
\frac{2g^2}{3\pi^2}\int d k \frac{k^2}{E}
\left(\frac{\xi_-}{\omega_-}
+\frac{\xi_+}{\omega_+}
\right)=1
\label{mgap2}
\ee 
\be
\frac{2 g^2}{9\pi^2}\int d k k^2
\left(\frac{1}{\omega_-}
+ \frac{1}{\omega_+}
\right)=1
\label{scgap2}
\ee 
where, $\omega_{\pm}=\sqrt{\Delta^2+\xi_{\pm}^2}$; $\xi_{\pm}=E\pm\nu$ and
$E=\sqrt{\zbf k^2 +M^2}$. To solve the above equations numerically, we take $\nu$ as an input and obtain the values of mass gap $M$ and the superconducting
gap $\Delta$ so that both the equations are satisfied simultaneously. 
The chemical potential $\nu$ is related to the fermi momentum 
as $k_f=\sqrt{\nu^2-M^2}$.
The resulting mass gap and the superconducting gap are plotted in Fig. 5
and Fig 6 respectively. For the sake of comparison we have also plotted 
the mass gap without diquark condensate (case--I) in Fig 5,and, superconducting
gap without chiral gap (case--II) in Fig. 6. As may be noted, the mass gap
does not change very much through the inclusion of diquark condensates. The
superconducting gap however starts at a lower threshold and varies slowly
compared to the case of no mass gap. Both the curves however merge at the chiral
restoration point.

\begin{figure}
\begin{center}
\epsfig{file=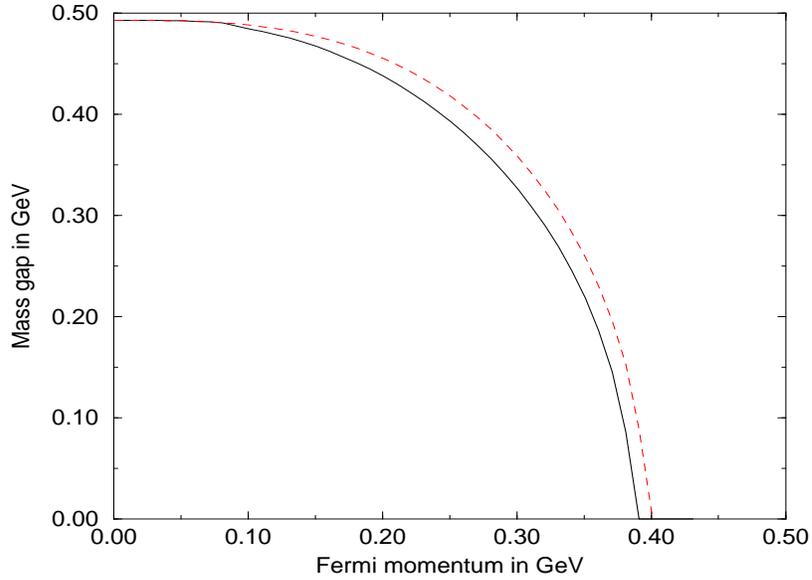,width=12 cm,height=9cm}
\end{center}
\caption{\em Mass gap $M$ as a function of fermi mometum. The dashed line
correspond to no diqurak condensates. The solid line corresponds to
both diquark and quark antiquark condensates.  }
\label{dmmgapfig}
\end{figure}

\begin{figure}
\begin{center}
\epsfig{file=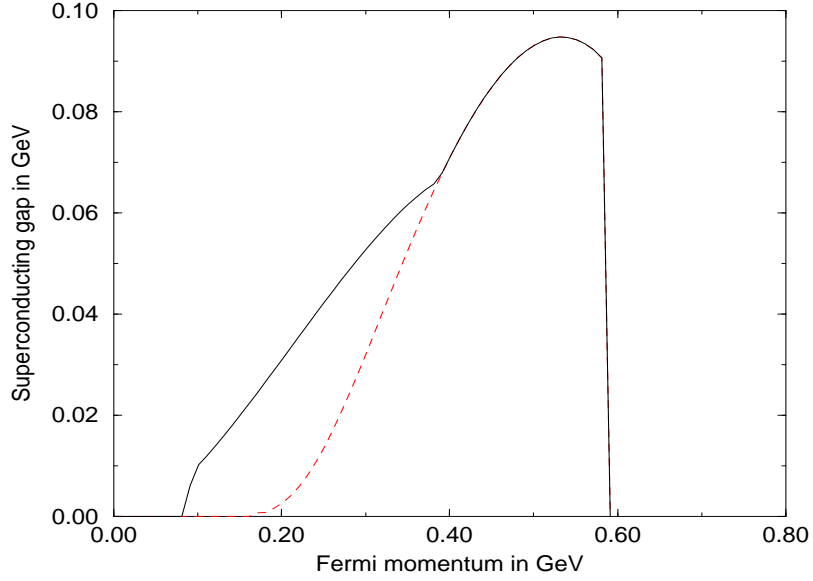,width=12 cm,height=9cm}
\end{center}
\caption{\em Superconducting gap $\Delta$ as a function of fermi mometum . 
The solid line corresponds to both diquark and
quark antiquark condensates. The dashed line corresponds to only
diquark condensates.}
\label{dmsgapfig}
\end{figure}
 We have also plotted the equation of state , pressure as a function of
fermi momentum in Fig.\ref{dmprefig}. The equation of state does not 
change very much compared to the case with only quark antiquark condensates.
\begin{figure}
\begin{center}
\epsfig{file=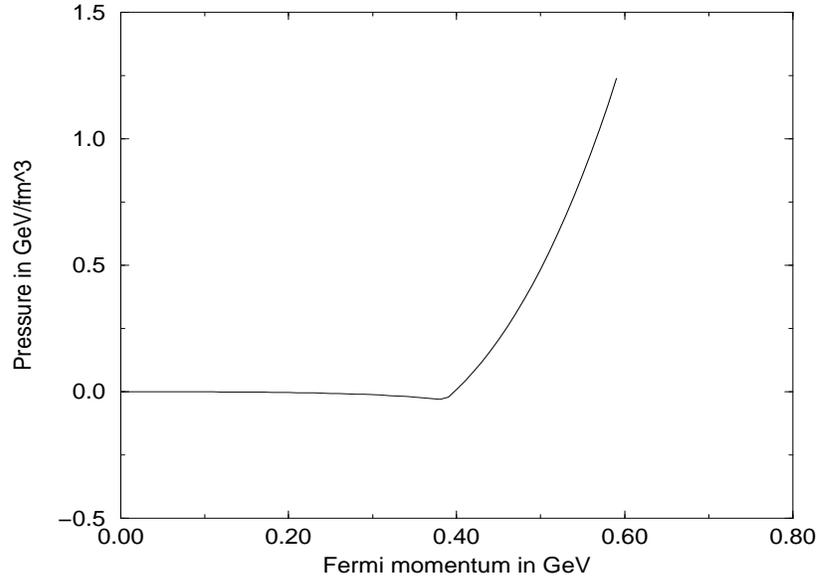,width=12 cm,height=9cm}
\end{center}
\caption{\em Pressure as a function of fermi momentum . }
\label{dmprefig}
\end{figure}
This should be expected as the effect of diquark condensate is small
in the region where  chiral condensate is nonvanishing. As may be evident
from Fig.[\ref{dmsgapfig}] there is a region from $k_f\simeq 0.1$ fm till
$k_f\simeq0.4$ fm where simultaneous existence of both types of condensates
is possible. However, in this region the pressure becomes negative.
As in case I, this will correspond to mechanical instability. The finite density
matter will break into droplets of finite density in which chiral
symmetry is restored and the matter is in a superconducting phase 
surrounded by empty space with zero pressure and density. 
Below $k_f=0.1$ fm the phase with only quark antiquark 
condensates and for $k_f>0.4$ fm the phase with diquark condensates is
thermodynamically feasible. However, one should not extrapolate to fermi momenta
higher than the cutoff momentum of the model to be consistent with
the philosophy of high momentum cutoff.
\end{itemize}
\subsection{The phase diagram}
To discuss the phase diagram we have solved the two gap equations
Eq.(\ref{mgap1}) and Eq. (\ref{scgap1}) at finite temperature and
density to calculate different thermodynamic quantities.
The result of such a calculation for the mass gap
is shown in Fig.(\ref{mgaptfig}). 
Similarly the superconducting gap
at different temperatures is shown in Fig.(\ref{scgaptfig}).
\begin{figure}
\begin{center}
\epsfig{file=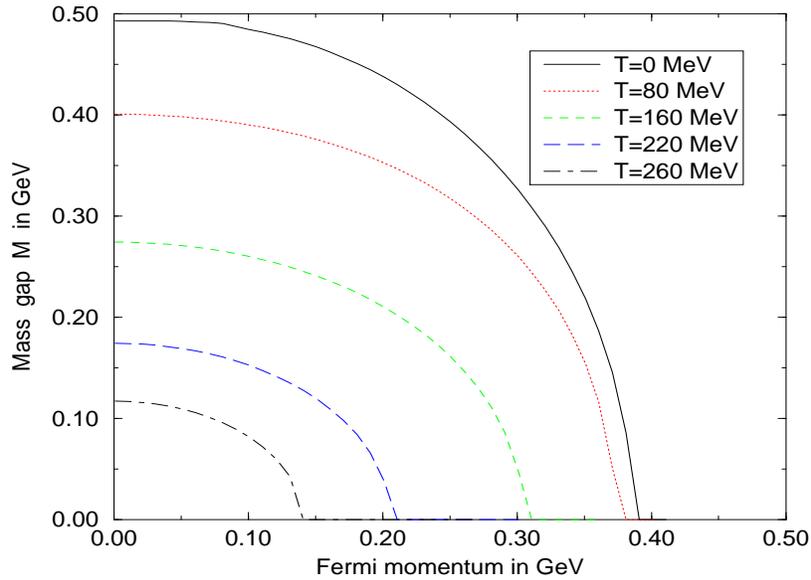,width=12 cm,height=9cm}
\end{center}
\caption{\em  Mass gap at different temperatures as a function
of fermi momentum.}
\label{mgaptfig}
\end{figure}

 The superconducting gap
decreases with temperature and vanishes at around 55 MeV. Similarly
the mass gap also decreases with temperature and vanishes at about 270 MeV
at zero density. The superconducting transition is a second order
phase transition. The chiral phase transition is first order at
zero density but is second order at high temperature. The pressure at 
different temperatures is shown in Fig.\ref{tricritfig} as a function of 
$(n/n_0)^{1/3}$, where n is number density and $n_0=(2/\pi^2)\nu_c^3$ is
the critical number density at zero temperature. The cusp in the pressure 
density curve, indicating a first order phase transition,
vanishes at about 84 MeV. At temperaures beyond that the pressure
increases monotonically with density.
\begin{figure}
\begin{center}
\epsfig{file=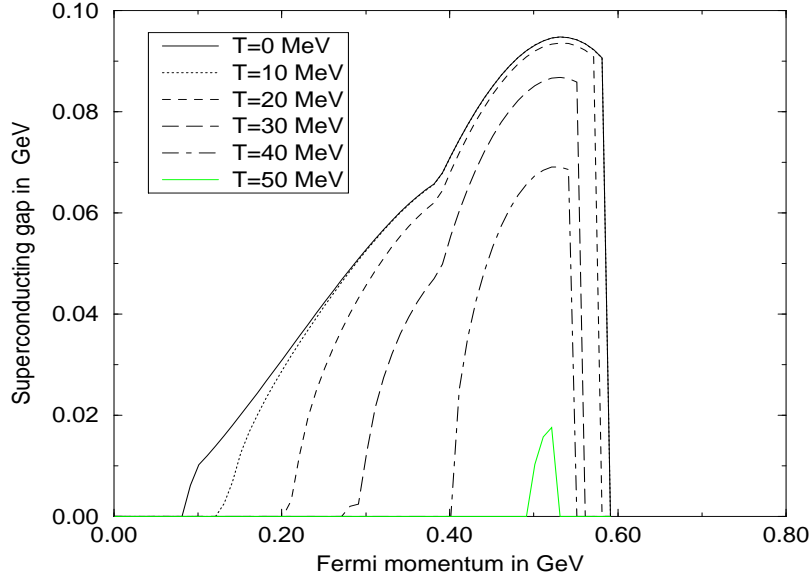,width=12 cm,height=9cm}
\end{center}
\caption{\em  Superconducting gap at different temperatures as a function
of fermi momentum.}
\label{scgaptfig}
\end{figure}
\begin{figure}
\begin{center}
\epsfig{file=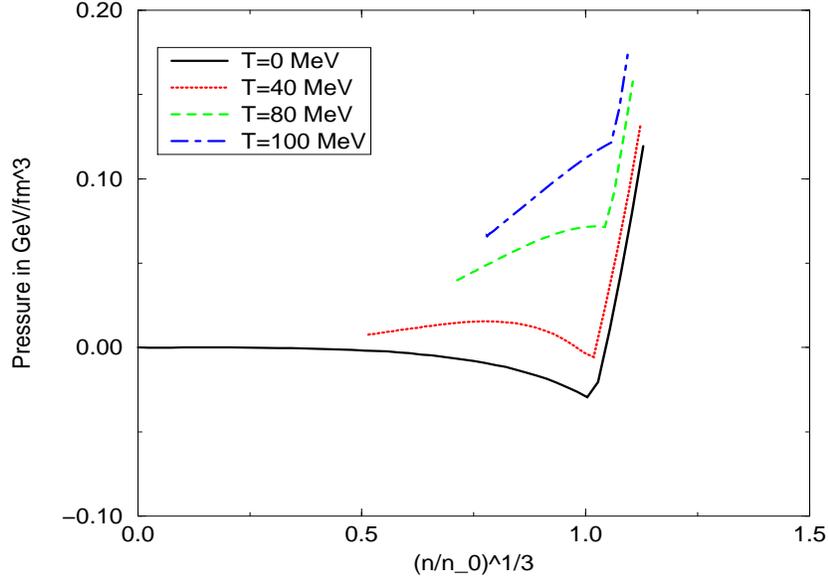,width=12 cm,height=9cm}
\end{center}
\caption{\em Pressure as a function of density at different
temperatures. The tricritical point turns out to be about 84 MeV.}
\label{tricritfig}
\end{figure}

To discuss the
critical line for the chiral transition
in the number density and temperature plane we need to solve the gap equation
(\ref{mgap1}) with mass gap equated to zero i.e. the critical line
satisfies the equation
\be
\frac{4 g^2}{3\pi^2}
\int_0^\Lambda dk k\left (1-\frac{1}{\exp(\beta(k-\nu))+1}
-\frac{1}{\exp(\beta(k+\nu))+1}\right )=1
\label{critm}
\ee

As the distribution functions are rapidly decreasing functions of momentum
 we may approximate the upper limit $\Lambda \rightarrow \infty$. In that
case Eq.(\ref {critm}) reduces to
\be
\nu^2+\tau^2=\nu_c^2
\label{nutau}
\ee
where,$\tau=\pi T /\sqrt{3}$ and
$\nu_c\equiv k_f^c=\Lambda(1-3\pi^2/(2g^2\Lambda^2)$,
the critical chemical potential or fermi momentum at zero temperature 
calculated earlierin Eq.(\ref{kfc}). It may be useful to write down 
this relation in terms
of number densities rather than the chemical potential using the relation
\be
n=\frac{6}{\pi^2}
\int_0^\Lambda dk k^2\left (\frac{1}{\exp(\beta(k-\nu))+1}
-\frac{1}{\exp(\beta(k+\nu))+1}\right )\approx\frac{2}{\pi^2}\nu(\nu^2+\pi^2T^2)
\label{ncrit}
\ee
where, we again take infinity
as the upper limit in the momentum integration as was done in Eq.(\ref{critm}).
This leads to the equation for the critical line in the number density plane as
\be
n/n_0=\left(1-(\frac{\tau}{\nu_c})^2\right)^{1/2}
\left(1+2(\frac{\tau}{\nu_c})^2\right)
\label{apcrit}
\ee
where, $n_0=(2/\pi^2)\nu_c^3$ is the critical density at zero temperature.
This approximate solution is shown by the dotted line in Fig{\ref{critfig}}.
The numerical solution for the critical line with the finite cutoff
for the momentum is shown by the solid curve in the same figure. As the 
chiral phase transition is a first order phase transition at
zero temperature and is second order at high temperature
there will be a tricritical point for the chiral phase transition 
below which there will be a mixed phase. This mixed phase here
will correspond to the phase with droplets of finite density 
quark matter in the superconducting but chiral symmetric phase surrounded
by empty space with $n=0, P=0,\Delta =0$ and chiral symmetry broken phase.
 This can be obtained by a Maxwell
construction to exclude the part of the phase diagram where pressure 
decreases with density. This is shown as the dashed curve in 
Fig.\ref{critfig}.  The tricritical point in this model turns out to be
84 MeV\cite{sarah,berges}.

 We have
also shown in Fig.{\ref{critfig}} the critical line for the color 
superconducting phase by the dot-dashed line. The decreasing part of this 
line is due to the fact that we have taken a cut off in the momentum and
at high enough density the superconducting gap vanishes as shown in Fig.
\ref{scgaptfig}. 

\begin{figure}
\begin{center}
\epsfig{file=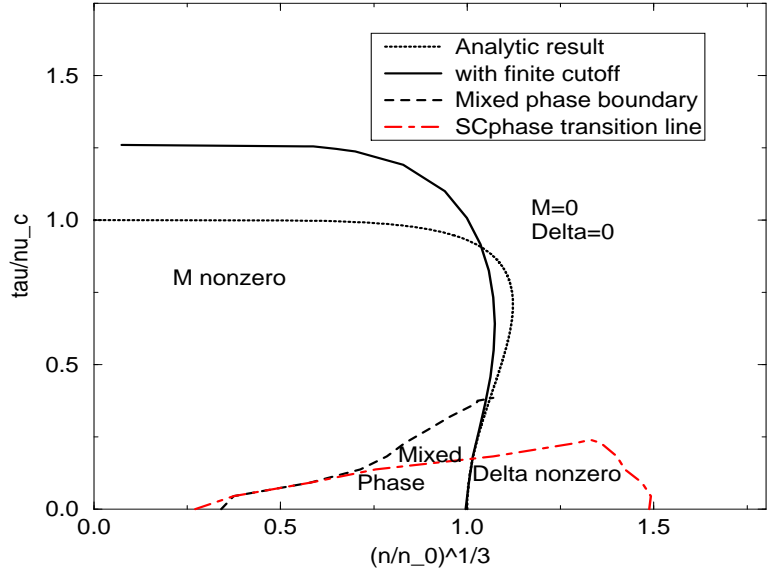,width=12 cm,height=9cm}
\end{center}
\caption{\em Phase diagram in density temperature plane}
\label{critfig}
\end{figure}
A detailed quantitative comparison of our results with those in 
Ref.\cite{berges} is not very meaningful since our model is a simple, schematic
one. However, it is very reassuring that the structure of the phase diagram 
is very similar to the one obtained by Berges and Rajagopal 
\cite{berges}.
\section{summary}
We have analysed in a current current point interaction 
model the structure of vacuum with quark antiquark as well as diquark
pairs. The methodology used is a variational one with an explicit
construct of the trial state. The present work is not based on a
mean field approximation
\cite{asakawa}.
Because of the point interaction structure we could solve 
for the gap functions explicitly. If we had taken the interaction term
with a potential the gap equation would become an integral equation. 
It will be interesting to include a realistic effective potential 
and solve for the gap equation. There appears to be a region 
depending upon the coupling where both
chiral condensates as well as diquark condensates are thermodynamically
feasible. Presence of diquark condensate does not modify the dynamical mass
of the quarks. The dynamical mass however affects the threshold for 
superconducting gap. The equation of state does not differ very much with
inclusion of diquark condensates.
We also obtain the complete phase diagram of the system in overall 
agreement with  that of Ref.\cite{berges}.
Finally, our results suggests that our simple model contains the 
essential physics of the system.

\acknowledgements HM would like to thank Amruta Mishra for numerous discussions.

\def \ltg{R.P. Feynman, Nucl. Phys. B 188, 479 (1981); 
K.G. Wilson, Phys. Rev. \zbf  D10, 2445 (1974); J.B. Kogut,
Rev. Mod. Phys. \zbf  51, 659 (1979); ibid  \zbf 55, 775 (1983);
M. Creutz, Phys. Rev. Lett. 45, 313 (1980); ibid Phys. Rev. D21, 2308
(1980); T. Celik, J. Engels and H. Satz, Phys. Lett. B129, 323 (1983)}

\def\berges {J. Berges, K. Rajagopal, Nucl. Phys. B538, 215, (1999).}
\def \svz {M.A. Shifman, A.I. Vainshtein and V.I. Zakharov,
Nucl. Phys. B147, 385, 448 and 519 (1979);
R.A. Bertlmann, Acta Physica Austriaca 53, 305 (1981)}

\def \spmbst {S.P. Misra, Phys. Rev. D35, 2607 (1987)}

\def \hmgrnv { H. Mishra, S.P. Misra and A. Mishra,
Int. J. Mod. Phys. A3, 2331 (1988)}

\def \snss {A. Mishra, H. Mishra, S.P. Misra
and S.N. Nayak, Phys. Lett 251B, 541 (1990)}

\def \amqcd { A. Mishra, H. Mishra, S.P. Misra and S.N. Nayak,
Pramana (J. of Phys.) 37, 59 (1991). }
\def\qcdtb{A. Mishra, H. Mishra, S.P. Misra 
and S.N. Nayak, Z.  Phys. C 57, 233 (1993); A. Mishra, H. Mishra
and S.P. Misra, Z. Phys. C 58, 405 (1993)}

\def \spmtlk {S.P. Misra, Talk on {\it `Phase transitions in quantum field
theory'} in the Symposium on Statistical Mechanics and Quantum field theory, 
Calcutta, January, 1992, hep-ph/9212287}

\def \hmnj {H. Mishra and S.P. Misra, Phys. Rev. D 48, 5376 (1993)}

\def \hmqcd {A. Mishra, H. Mishra, V. Sheel, S.P. Misra and P.K. Panda,
hep-ph/9404255 (1994)}

\def \amcrl {A. Mishra, H. Mishra and S.P. Misra, Z. Phys. C 57, 241 (1993)}

\def \higgs { S.P. Misra, in {\it Phenomenology in Standard Model and Beyond}, 
Proceedings of the Workshop on High Energy Physics Phenomenology, Bombay,
edited by D.P. Roy and P. Roy (World Scientific, Singapore, 1989), p.346;
A. Mishra, H. Mishra, S.P. Misra and S.N. Nayak, Phys. Rev. D44, 110 (1991)}

\def \nmtr {A. Mishra, 
H. Mishra and S.P. Misra, Int. J. Mod. Phys. A5, 3391 (1990); H. Mishra,
 S.P. Misra, P.K. Panda and B.K. Parida, Int. J. Mod. Phys. E 1, 405, (1992);
 {\it ibid}, E 2, 547 (1993); A. Mishra, P.K. Panda, S. Schrum, J. Reinhardt
and W. Greiner, to appear in Phys. Rev. C}

\def \dtrn {P.K. Panda, R. Sahu and S.P. Misra, 
Phys. Rev C45, 2079 (1992)}

\def \qcd {G. K. Savvidy, Phys. Lett. 71B, 133 (1977);
S. G. Matinyan and G. K. Savvidy, Nucl. Phys. B134, 539 (1978); N. K. Nielsen
and P. Olesen, Nucl.  Phys. B144, 376 (1978); T. H. Hansson, K. Johnson,
C. Peterson Phys. Rev. D26, 2069 (1982)}

\def \cornwal {J.M. Cornwall, Phys. Rev. D26, 1453 (1982)}

\def \mndglv {J. E. Mandula and M. Ogilvie, Phys. Lett. 185B, 127 (1987)}

\def \schwinger {J. Schwinger, Phys. Rev. 125, 1043 (1962); ibid,
127, 324 (1962)}

\def \schutte {D. Schutte, Phys. Rev. D31, 810 (1985)}

\def \amspm {A. Mishra and S.P. Misra, Z. Phys. C 58, 325 (1993)}

\def \gft{ For gauge fields in general, see e.g. E.S. Abers and 
B.W. Lee, Phys. Rep. 9C, 1 (1973)}

\def \gribov {V.N. Gribov, Nucl. Phys. B139, 1 (1978)}

\def \spm78 {S.P. Misra, Phys. Rev. D18, 1661 (1978); {\it ibid}
D18, 1673 (1978)} 

\def \lopr {A. Le Youanc, L.  Oliver, S. Ono, O. Pene and J.C. Raynal, 
Phys. Rev. Lett. 54, 506 (1985)}

\def \spphi {S.P. Misra and S. Panda, Pramana (J. Phys.) 27, 523 (1986);
S.P. Misra, {\it Proceedings of the Second Asia-Pacific Physics Conference},
edited by S. Chandrasekhar (World Scientfic, 1987) p. 369}

\def\spmdif {S.P. Misra and L. Maharana, Phys. Rev. D18, 4103 (1978); 
    S.P. Misra, A.R. Panda and B.K. Parida, Phys. Rev. Lett. 45, 322 (1980);
    S.P. Misra, A.R. Panda and B.K. Parida, Phys. Rev. D22, 1574 (1980)}

\def \spmvdm {S.P. Misra and L. Maharana, Phys. Rev. D18, 4018 (1978);
     S.P. Misra, L. Maharana and A.R. Panda, Phys. Rev. D22, 2744 (1980);
     L. Maharana,  S.P. Misra and A.R. Panda, Phys. Rev. D26, 1175 (1982)}

\def\spmthr {K. Biswal and S.P. Misra, Phys. Rev. D26, 3020 (1982);
               S.P. Misra, Phys. Rev. D28, 1169 (1983)}

\def \spmstr { S.P. Misra, Phys. Rev. D21, 1231 (1980)} 

\def \spmjet {S.P. Misra, A.R. Panda and B.K. Parida, Phys. Rev Lett. 
45, 322 (1980); S.P. Misra and A.R. Panda, Phys. Rev. D21, 3094 (1980);
  S.P. Misra, A.R. Panda and B.K. Parida, Phys. Rev. D23, 742 (1981);
  {\it ibid} D25, 2925 (1982)}

\def \arpftm {L. Maharana, A. Nath and A.R. Panda, Mod. Phys. Lett. 7, 
2275 (1992)}

\def \van {R. Van Royen and V.F. Weisskopf, Nuov. Cim. 51A, 617 (1965)}

\def \rchpi {S.R. Amendolia {\it et al}, Nucl. Phys. B277, 168 (1986)}

\def \chrl{ Y. Nambu, Phys. Rev. Lett. \zbf 4, 380 (1960);
A. Amer, A. Le Yaouanc, L. Oliver, O. Pene and
J.C. Raynal, Phys. Rev. Lett.\zbf  50, 87 (1983);
ibid, Phys. Rev.\zbf  D28, 1530 (1983); 
M.G. Mitchard, A.C. Davis and A.J.
Macfarlane, Nucl. Phys. \zbf B325, 470 (1989);
B. Haeri and M.B. Haeri, Phys. Rev.\zbf  D43,
3732 (1991);; V. Bernard, Phys. Rev.\zbf  D34, 1601 (1986);
 S. Schram and
W. Greiner, Int. J. Mod. Phys. \zbf E1, 73 (1992)}
\def\finger{ J.R. Finger and J.E. Mandula, Nucl. Phys. \zbf B199, 168 (1982).}
\def\davis{S.L. Adler and A.C. Davis, Nucl. Phys.\zbf  B244, 469 (1984) .}
\def\alkofer{R. Alkofer and P. A. Amundsen, Nucl. Phys.\zbf B306, 305 (1988)}
\def\klev{S.P. Klevensky, Rev. Mod. Phys.\zbf  64, 649 (1992);}
 \def\bhalerao{S. Li, R.S. Bhalerao and R.K. Bhaduri, Int. J. Mod. Phys. 
 \zbf A6, 501 (1991)}
\def\asakawa{M. Asakawa and K. Yazaki, Nucl. Phys. A504,668 (1989).}

\def \spmijp { S.P. Misra, Ind. J. Phys. 61B, 287 (1987)}

\def \feynman {R.P. Feynman and A.R. Hibbs, {\it Quantum mechanics and
path integrals}, McGraw Hill, New York (1965)}

\def \glstn{ J. Goldstone, Nuov. Cim. \zbf 19, 154 (1961);
J. Goldstone, A. Salam and S. Weinberg, Phys. Rev. \zbf  127,
965 (1962)}

\def \anderson {P.W. Anderson, Phys. Rev. \zbf {110}, 827 (1958)}

\def \nambu{ Y. Nambu, Phys. Rev. Lett. \zbf 4, 380 (1960)}

\def\donogh {J.F. Donoghue, E. Golowich and B.R. Holstein, {\it Dynamics
of the Standard Model}, Cambridge University Press (1992)}

\def\satz {T. Matsui and H. Satz, Phys. Lett. B178, 416 (1986)}

\def\cps {C. P. Singh, Phys. Rep. 236, 149 (1993)}

\def\prliop {A. Mishra, H. Mishra, S.P. Misra, P.K. Panda and Varun
Sheel, Int. J. of Mod. Phys. E 5, 93 (1996)}

\def\hmcor {V. Sheel, H. Mishra and J.C. Parikh, Phys. Lett. B382, 173
(1996); {\it biid}, to appear in Int. J. of Mod. Phys. E}
\def\cort { V. Sheel, H. Mishra and J.C. Parikh, Phys. ReV D59,034501 (1999);
{\it ibid}Prog. Theor. Phys. Suppl.,129,137, (1997).}
\def\amph4{Amruta Mishra and Hiranmaya Mishra, J. Phys. G23,143, (1997).}

\def\surcor {E.V. Shuryak, Rev. Mod. Phys. 65, 1 (1993)} 

\def\stevenson {A.C. Mattingly and P.M. Stevenson, Phys. Rev. Lett. 69,
1320 (1992); Phys. Rev. D 49, 437 (1994)}

\def\mac {M. G. Mitchard, A. C. Davis and A. J. Macfarlane,
 Nucl. Phys. B 325, 470 (1989)} 
\def\tfd
 {H.~Umezawa, H.~Matsumoto and M.~Tachiki {\it Thermofield dynamics
and condensed states} (North Holland, Amsterdam, 1982) ;
P.A.~Henning, Phys.~Rep.253, 235 (1995).}

\def \neglecor{M.-C. Chu, J. M. Grandy, S. Huang and 
J. W. Negele, Phys. Rev. D48, 3340 (1993);
ibid, Phys. Rev. D49, 6039 (1994)}

\def\revdata {Particle Data Group, Phys. Rev. D 50, 1173 (1994)}

\def\sinp {S.P. Misra, Indian J. Phys., {\bf 70A}, 355 (1996)}

\def\bryman {D.A. Bryman, P. Deppomier and C. Le Roy, Phys. Rep. 88,
151 (1982)}

\def\thooft {G. 't Hooft, Phys. Rev. D 14, 3432 (1976); D 18, 2199 (1978);
S. Klimt, M. Lutz, U. Vogl and W. Weise, Nucl. Phys. A 516, 429 (1990)}
\def\alkz { R. Alkofer, P. A. Amundsen and K. Langfeld, Z. Phys. C 42,
199(1989), A.C. Davis and A.M. Matheson, Nucl. Phys. B246, 203 (1984).}
\def\sarah {T.M. Schwartz, S.P. Klevansky, G. Papp, Phys. Rev. C   (1999)}
\def\wil{M. Alford, K.Rajagopal, F. Wilczek, Phys. Lett. B422,247(1998)i
{\it{ibid}}Nucl. Phys. B537,443 (1999).}
\def\sursc{R.Rapp, T.Schaefer, E. Shuryak and M. Velkovsky Phys. Rev. Lett.
81, 53(1998),{\it {ibid}},hep-ph/9904353.}
\def\pisarski{
D. Bailin and A. Love, Phys. Rep. 107 (1984) 325,
D. Son, Phys. Rev. D59 (1999) 094019,
T. Schaefer and F. Wilczek, Phys. Rev. D60 (1999) 114033,
D. Rischke and R. Pisarski, Phys. Rev. D61 (2000) 051501,
D. K. Hong, V. A. Miransky, I. A. Shovkovy, L.C. Wiejewardhana, 
Phys. Rev. D61 (2000) 056001.}
\def\leblac {M. Le Bellac, {\it Thermal Field Theory}(Cambridge, Cambridge University
Press, 1996).}
\def\bcs{A.L. Fetter and J.D. Walecka, {\it Quantum Theory of Many
particle Systems} (McGraw-Hill, New York, 1971).}
\def\alexander{Aleksander Kocic, Phys. Rev. D33, 1785,(1986).}


\begin{references}

\bibitem{sur}
 E.V.~Shuryak, {\it The QCD vacuum, hadrons and the superdense matter} 
(World Scientific, Singapore, 1988).

\bibitem{svz} 
M.A. Shifman, A.I. Vainshtein and V.I. Zakharov, Nucl.Phys. B147, 
 385, 448 and 519(1979).

\bibitem{chrl} \chrl
\bibitem{finger} \finger
\bibitem{davis} \davis
\bibitem{alkofer} \alkofer
\bibitem{klev} \klev
\bibitem{bhalerao}\bhalerao
\bibitem{amcrl} \amcrl
\bibitem{hmnj} \hmnj
\bibitem{alkz} \alkz
\bibitem{wil} \wil
\bibitem{sursc} \sursc
\bibitem{pisarski} \pisarski
\bibitem{sarah} \sarah
\bibitem{berges}\berges
\bibitem{tfd}\tfd
\bibitem{spmtlk} \spmtlk
\bibitem{amspm} \amspm
\bibitem{spm78} \spm78
\bibitem{lopr} \lopr
\bibitem{sinp} \sinp
\bibitem{bcs}\bcs
\bibitem{amph4}\amph4
\bibitem{cort}\cort
\bibitem{qcdtb} \qcdtb
\bibitem{leblac}\leblac
\bibitem{asakawa}\asakawa
\bibitem{alexander}\alexander
\vfil
\end{references}
\end{document}